# LIGHTCURVES FOR ASTEROIDS 18301 KONYUKHOV AND 2022 WEST

Sinhué A.R. Haro-Corzo, Luis A. Villegas
Escuela Nacional de Estudios Superiores Unidad Morelia
Universidad Nacional Autónoma de México
Morelia, Michoacán, MÉXICO
sharo@enesmorelia.unam.mx

Lorenzo Olguín, Julio C. Saucedo, María E. Contreras
Departamento de Investigación en Física
Universidad de Sonora
Hermosillo, Sonora, MÉXICO

Pedro V. Sada
Departamento de Física y Matemáticas
Universidad de Monterrey
Av. I. Morones Prieto 4500 Pte.
San Pedro Garza García, N.L. 66238
MÉXICO

Sandra A. Ayala, Jaime R. Garza
Facultad de Ciencias Físico-Matemáticas
Universidad Autónoma de Nuevo León
Monterrey, Nuevo León
MÉXICO

Juan Segura-Sosa
Facultad de Ciencias Físico-Matemáticas
Universidad Autónoma de Coahuila
Saltillo, Coahuila, MÉXICO

Claudia P. Benítez-Benítez
Escuela Superior de Física y Matemáticas
Instituto Politécnico Nacional
Cd. de México, MÉXICO

We report photometric analysis of two main-belt asteroids observed at the Observatorio Astronomico Nacional in the Sierra San Pedro Martir, Baja California, México. For 18301 Konyukhov our derived intrinsic rotation period is 2.6667±0.0003 h with an amplitude of 0.16 magnitudes. To the best of our knowledge, this is the first lightcurve reported for this asteroid. In the case of 2022 West our derived intrinsic rotation period is 14.1385 ±0.0031 h with an amplitude of 0.54 magnitudes.

The Mexican effort to achieve coordinated simultaneous photometric observations of asteroids is embodied by the Mexican Asteroid Photometric Campaign (hereafter CMFA). Since 2015, more than 10 asteroids have been photometrically observed and analyzed (Sada et al 2016, 2017, 2018). In this work we present photometric data of two main-belt asteroids supplemental to the 2016 CMFA. These objects were observed in the period 2016 August-November as  targets of opportunity at the Observatorio Astronómico Nacional at San Pedro Mártir (hereafter OAN-SPM; MPC 679) in Baja California, México. The observations were carried out with the 0.84-m f/15 Ritchey-Chretien telescope. We used a 2048×2048 pix$^2$ E2V-4240 cryogenic CCD operating at a temperature of -110 °C. The images were generally binned 2 × 2 with a final field of view of  7.6 × 7.6 arcmin$^2$. All observations were unfiltered. The observed images were corrected using standard IRAF routines in order to correct them for nightly bias, dark current and flat-field effects. We used MPO Canopus (V9.5.0.14, BDW Publishing, 2017) to carry out differential photometric measurements and lightcurve analysis.

18301 Konyukhov (1979 QZ9) was discovered on 1979  August 27 by  N. S. Chernykh and was named after the Russian traveler Fyodor Fyodorovich Konyukhov. It is an outer main-belt asteroid with H magnitude of 13.4 (see JPL Horizons webpage). The asteroid  was observed at the OAN-SPM on six nights of 2016 (Aug 20, 21, 22 &  23; Oct 15 & 18). During analysis, Aug 22 observations were treated as two separate sessions because the asteroid passed in front of a bright star. A total of 395 data points were used to construct its lightcurve. From this curve, we derived an intrinsic rotation period of 2.6667±0.0003 h with an amplitude of ~0.17 mag (see Figure for 18301 Konyukhov). This period is in the range of a typical asteroid in the main-belt.  According to the JPL Small-Body Database there are a total of 1738 observations (all types) since 1979. However, as far as we know there is not photometry nor lightcurve data reported in the literature.

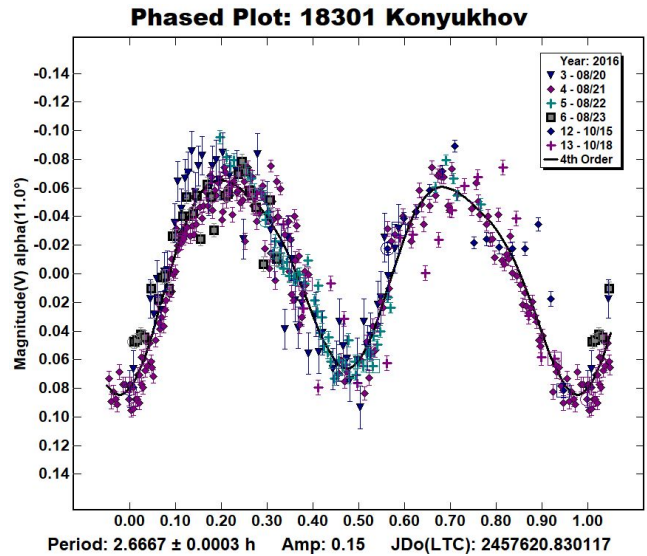

2022 West (1938 CK) was discovered on 1938 February 7 by K. Reinmuth.  It is a  main-belt asteroid with  H magnitude 11.6 (see JPL Horizons webpage). 2022 West was reported by Franco and Marchini (2017) to have a rotation period of 14.14±0.01 h. In this work 2022 West was observed on seven nights of 2016 (Aug 20 & 21; Oct 11,15 & 18; Nov 10 & 11). During analysis, Oct 18 observations were treated as two separate sessions because two different sets of bright and comparison stars were used. A total of 287 data points were used to construct its lightcurve. Based on this curve, we derived an intrinsic rotation period of 14.1385±0.0031 h with an amplitude of 0.54 mag (see Figure for 2022 West). In this case, the period is 3.2 times more precise than previous works because this asteroid was observed over a longer period of time.

| Number | Name | 2016 mm/dd | Pts | Phase | $L_{PAB}$ | $B_{PAB}$ | Period(h) | P.E. | Amp | A.E. | Grp |
|---|---|---|---|---|---|---|---|---|---|---|---|
| 18301 | Konyukhov | 08/20-10/18 | 395 | 10.9,12.7 | 312.4 | 0.69 | 2.6667 | 0.0003 | 0.15 | 0.02 | MB |
| 2022 | West | 08/20-11/11 | 287 | 22.5, 10.2 | 27.9 | 2.74 | 14.1385 | 0.0031 | 0.54 | 0.04 | MB |

Table I. Observing circumstances and results. Pts is the number of data points. The phase angle is given for the first and last date. $L_{PAB}$ and $B_{PAB}$ are the approximate phase angle bisector longitude and latitude at mid-date range (see Harris *et al.*, 1984) which values were extracted from https://ssd.jpl.nasa.gov/horizons.cgi#top . Grp is the asteroid family/group (Warner *et al.*, 2009).

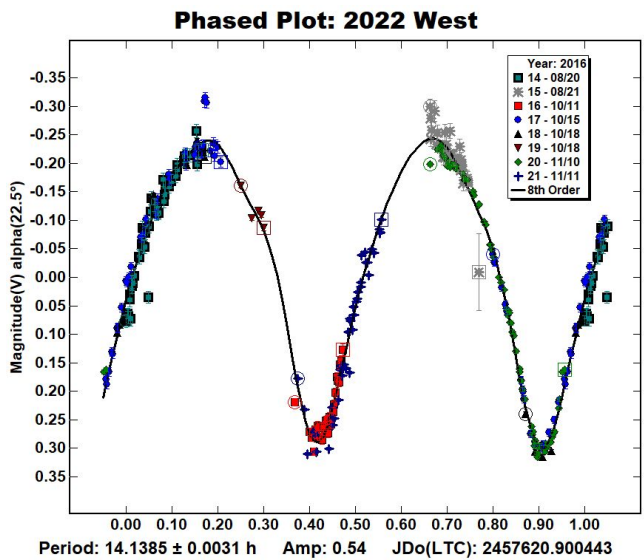

## Acknowledgements

The results presented in this report are based upon observations carried out at the Observatorio Astronómico Nacional on the Sierra San Pedro Mártir (OAN-SPM), Baja California, México. IRAF is distributed by the National Optical Astronomy Observatory, which is operated by the Association of Universities for Research in Astronomy (AURA) under a cooperative agreement with the National Science Foundation.